\documentclass[11pt]{article}
\usepackage{amssymb}
\usepackage[perpage,symbol]{footmisc}
\usepackage{listings}
\usepackage{amsfonts}
\usepackage{enumerate}
\usepackage{amsmath}
\usepackage{cite}
\usepackage{array}
\usepackage{booktabs}

\begin{document}

\newtheorem{theorem}{Theorem}[section]
\newtheorem{corollary}[theorem]{Corollary}
\newtheorem{definition}[theorem]{Definition}
\newtheorem{proposition}[theorem]{Proposition}
\newtheorem{lemma}[theorem]{Lemma}
\newtheorem{example}[theorem]{Example}
\newenvironment{proof}{\noindent {\bf Proof.}}{\rule{3mm}{3mm}\par\medskip}
\newcommand{\remark}{\medskip\par\noindent {\bf Remark.~~}}
\title{Codes from Goppa codes}
\author{Chunlei Liu\footnote{Shanghai Jiao Tong Univ., Shanghai 200240, clliu@sjtu.edu.cn.}}
\date{}
\maketitle
\thispagestyle{empty}

\abstract{The Fr\"{o}benius acts on a Goppa code whose structure polynomial has coefficients in the symbol field. Its fixed codewords form a subcode. Deleting the redundance caused by repetition,  we obtain a new code. We call this code a reduced Goppa code. It is proved that, as classical Goppa codes, the reduced Goppa codes approach to the Gilbert-Varshamov bound. It is also proved that, if the characteristic of the symbol field is small, the decoding time for a reduced Goppa code of codeword length $n$ is $O(n^2\log^{\nu}n)$ in operations of the symbol field. Here the constant $\nu$ is determined by the performance of polynomial factorisation algorithms. Recall that the decoding time for classical Goppa codes of codeword length $n$ is $O(n^2\log^{\nu}n)$ in operations of a field whose size is about the size of the set of locations. By comparison, reduced Goppa codes can be decoded much quicker that classical Goppa codes.}


\section{\small{INTRODUCTION}}
\hskip .2in Reed-Solomon codes \cite{RS}, BCH codes \cite{BC}, \cite{Ho} and Goppa codes \cite{Go} are widely used error correcting codes due to their fast decoding algorithms based on the Berlekamp-Massey algorithm \cite{Be68,Ma}. We shall construct new codes from Goppa codes. These new codes will have a faster decoding algorithm and have performances competent to the classical Goppa codes.

We recall the notion of Goppa codes in a sense as narrow as possible. As any linear code, a Goppa code has a Galois field ${\rm GF}(q)$ as its symbol field. To ensure quick decoding, we require $q$ to be a power of a small prime number. Its location field is an extension of ${\rm GF}(q)$, say ${\rm GF}(q^m)$. Correspondingly, its set of locations is ${\mathbb Z}/(q^m-1)$, which is canonically isomorphic to the multiplicative group ${\rm GF}(q^m)^{\times}$. To ensure a lower bound for its minimal distance, it has a designed distance $\delta$. To get a code of good performance, it has a structure polynomial $g$ of degree $\delta-1$ having no roots in ${\rm GF}(q^m)^{\times}$. For simplicity, we require $g$ to have coefficients in ${\rm GF}(q)$. The Goppa code with these parameters is denoted as $\Gamma(g,\delta,m,q)$.
Its codewords satisfy the congruence $$\sum_{i=0}^{q^m-2}\frac{c_i}{x-\beta^i}\equiv0({\rm mod} g(x)),$$
where $\beta$ is a fixed primitive element of ${\rm GF}(q^m)$.

The Fr\"{o}benius of the Galois group of ${\rm GF}(q^m)$ over ${\rm GF}(q)$ acts on $\Gamma(g,\delta,m,q)$. Its fixed codewords form a subcode $\Gamma^{\rm F}(g,\delta,m,q)$. Deleting the redundance caused by repetition,  we obtain a new code ${\rm R}\Gamma(g,\delta,m,q)$. We call this code a reduced Goppa code. The precise construction of reduced Goppa codes is given in Definition \ref{construction}. It is proved that the reduced Goppa codes approach to the Gilbert-Varshamov bound (see Theorem \ref{GV-bound}). It is also proved that the decoding time for a reduced Goppa code of codeword length $n$ is $O(n^2\log^{\nu}n)$ in $q$-ary operations (see Theorem \ref{decoding-time}).  Here the constant $\nu$ is determined by the performance of polynomial factorisation algorithms. Recall that decoding the classical Goppa codes of codeword length $n=q^m-1$ requires $O(n^2\log^{\nu}n)$ in $q^m$-ary operations. By comparison, reduced Goppa codes can be decoded much quicker than classical Goppa codes.

{\bf Acknowledgement.}
The author thanks Jiyou Li for pointing out some mistakes.
\section{Construction}
\hskip 0.1in In this section we introduce the action of the Fr\"{o}benius on Goppa codes, and give the construction of reduced Goppa codes in Definition \ref{construction}.

\begin{definition}The Fr\"{o}benius ${\rm Frob}$ on the group ${\mathbb Z}/(q^m-1)$ is a permutation on ${\mathbb Z}/(q^m-1)$ defined by the formula:
$${\rm Frob}(i+(q^m-1){\mathbb Z})=qi+(q^m-1){\mathbb Z}.$$\end{definition}\begin{definition}The Fr\"{o}benius ${\rm Frob}$ on the space ${\rm GF}(q)^{{\mathbb Z}/(q^m-1)}$ is a permutation on ${\rm GF}(q)^{{\mathbb Z}/(q^m-1)}$ defined by the formula:
$${\rm Frob}(r)_i=r_{q^{m-1}i}.$$\end{definition}\begin{lemma}The Goppa code $\Gamma(g,\delta,m,q)$ is stable under the Fr\"{o}benius.\end{lemma}
\begin{proof} Let $c\in\Gamma(g,\delta,m,q)$.
That is
$$\sum_{i=0}^{q^m-2}\frac{c_{i}}{x-\beta^{i}}\equiv0
({\rm mod} g(x)).$$
As $g$ has coefficients in ${\rm GF}(q)$, we infer
$$\sum_{i=0}^{q^m-2}\frac{c_{i}}{x-\beta^{qi}}
\equiv0({\rm mod} g(x)).$$
Then
\begin{eqnarray*}
  \sum_{i=0}^{q^m-2}\frac{{\rm Frob}(c)_i}{x-\beta^i}
&=&
\sum_{i=0}^{q^m-2}\frac{c_{q^{m-1}i}}{x-\beta^i}
({\rm mod} g(x)) \\
&=&
\sum_{i=0}^{q^m-2}\frac{c_{i}}{x-\beta^{qi}}
({\rm mod} g(x))\\
&\equiv&0({\rm mod} g(x)). \\
\end{eqnarray*}
That is, ${\rm Frob}(c)\in \Gamma(g,\delta,m,q)$.
The lemma is proved.
\end{proof}\begin{definition}The subcode of ${\Gamma}(g,\delta,m,q)$ formed of Fr\"{o}benius fixed codewords is the code
$$\Gamma^{\rm F}(g,\delta,m,q)=\{c\in {\Gamma}(g,\delta,m,q)\mid {\rm Frob}(c)=c\}.$$\end{definition}
\begin{definition}The set of orbits of ${\mathbb Z}/(q^m-1)$ under the Fr\"{o}benius is denoted as $O(m,q)$.\end{definition}
\begin{definition}Let $c\in \Gamma^{\rm F}(g,\delta,m,q)\}$. For $o\in O(m,q)$, we define
$$c_o=c_i,\ \forall i\in o.$$\end{definition}
\begin{definition}\label{construction}The code
$${\rm R}\Gamma(g,\delta,m,q)=\{(c_o)_{o\in O(m,q)}\mid c\in \Gamma^{\rm F}(g,\delta,m,q)\}$$ is called a reduced Goppa code.\end{definition} We see that ${\rm R}\Gamma(g,\delta,m,q)$ is obtained from ${\Gamma}^{\rm F}(g,\delta,m,q)$ by deleting redundance caused by repetition, and, as a space over
${\rm GF}(q)$, is canonically isomorphic to $\Gamma^{\rm F}(g,\delta,m,q)$.
\section{Lower bounds on minimal distance and dimension}
\hskip 0.1in In this section we give lower bounds for the minimal distance and dimension of reduced Goppa codes. The main results are Theorems \ref{distance-bound} and \ref{dimension-bound}.
\begin{theorem}[distance bound]\label{distance-bound}
$${\rm dist}{\rm R}\Gamma(g,\delta,m,q)\geq\frac{1}{m}{\rm dist}\Gamma^{\rm F}(g,\delta,m,q)\geq\frac{\delta}{m}.$$\end{theorem}
\begin{proof} Let $c\in{\rm R}\Gamma(g,\delta,m,q)$ be of minimal distance $d$. For $i\in{\mathbb Z}/(q^m-1)$,
define
$$c_i=c_o, \ i\in o.$$ Then $(c_i)_{i\in{\mathbb Z}/(q^m-1)}\in \Gamma^{\rm F}(g,\delta,m,q)$, and
$$w(c_i)_{i\in{\mathbb Z}/(q^m-1)}\leq md.$$
It follows that
$${\rm dist}\Gamma^{\rm F}(g,\delta,m,q)\leq md\leq m\cdot{\rm dist}{\rm R}\Gamma(g,\delta,m,q).$$
The theorem now follows.\end{proof}
\begin{lemma}If $g(x)=\sum_{i=0}^{\delta-1}g_ix^i$, then the polynomials
$$\sum_{i=1}^{\delta-1-j}g_{i+j} x^i,\ j=0,1,\cdots,\delta-2$$
are linearly independent in ${\rm GF}(q)[x]/(g(x))$.\end{lemma}
\begin{proof}
This follows from the fact that
$(\sum_{i=1}^{\delta-1-j}g_{i+j} x^i)_{j=0,1,\cdots,\delta-2}$ is equal to
$$(x^{\delta-1},x^{\delta-2},\cdots,x)
\left(
     \begin{array}{cccc}
       g_{\delta-1} & 0 & 0 & 0 \\
       g_{\delta-2} & g_{\delta-1} & 0 & 0 \\
       \vdots & \vdots & \ddots & 0 \\
       g_1 & g_2 & \cdots & g_{\delta-1} \\
     \end{array}
   \right)$$
\end{proof}
\begin{lemma}[codeword equations]$c\in \Gamma(g,\delta,m,q)$ if and only if
$$\sum_{i=0}^{q^m-2}\frac{c_i}{g(\beta^i)}\beta^{ij}=0,\ j=0,1,\cdots,\delta-2.$$\end{lemma}

\begin{proof} We have
\begin{eqnarray*}
  \sum_{i=0}^{q^m-2}\frac{g(x)-g(\beta^i)}{g(\beta^i)}\frac{c_i}{x-\beta^i}
&=& \sum_{i=0}^{q^m-2}\frac{c_i}{g(\beta^i)}\sum_{u=1}^{\delta-1}g_u
\frac{x^u-\beta^{iu}}{x-\beta^i}\\&=& \sum_{i=0}^{q^m-2}\frac{c_i}{g(\beta^i)}\sum_{u=1}^{\delta-1}g_u
\sum_{j=0}^{u-1}x^{u-j}\beta^{ij}\\&=& \sum_{i=0}^{q^m-2}\frac{c_i}{g(\beta^i)}
\sum_{j=0}^{\delta-2}\beta^{ij}\sum_{u=j+1}^{\delta-1}g_u x^{u-j}\\&=& \sum_{j=0}^{\delta-2}\left(\sum_{i=0}^{q^m-2}c_i\frac{\beta^{ij}}{g(\beta^i)}
\right)\sum_{i=1}^{\delta-1-j}g_{i+j} x^i.
\end{eqnarray*}
By the last lemma, the polynomials
$$\sum_{i=1}^{\delta-1-j}g_{i+j} x^i,\ j=0,1,\cdots,\delta-2$$
are linearly independent.
It follows that
\begin{eqnarray*}
 && \sum_{i=0}^{q^m-2}\frac{c_i}{g(\beta^i)}\beta^{ij}=0,\ j=0,1,\cdots,\delta-2 \\
&\Leftrightarrow& \sum_{i=0}^{q^m-2}\frac{g(x)-g(\beta^i)}{g(\beta^i)}\frac{c_i}{x-\beta^i}\equiv
0({\rm mod} g(x)) \\
   &\Leftrightarrow& \sum_{i=0}^{q^m-2}\frac{c_i}{x-\beta^i}\equiv
0({\rm mod} g(x))\\
&\Leftrightarrow&c\in \Gamma(g,\delta,m,q).
\end{eqnarray*}
The lemma is proved.
\end{proof}
\begin{lemma}[codeword congruence]$c\in \Gamma(g,\delta,m,q)$
if and only if
$$\sum_{i=0}^{q^m-2}\frac{c_i}{g(\beta^{i})}\frac{1}{1-\beta^ix}\equiv0({\rm mod }x^{\delta-1}).$$\end{lemma}
\begin{proof}
We have
\begin{eqnarray*}
  \sum_{i=0}^{q^m-2}\frac{c_i}{g(\beta^{i})}\frac{1}{1-\beta^ix}&=&
 \sum_{i=0}^{q^m-2}\frac{c_i}{g(\beta^{i})}\sum_{j=0}^{+\infty}\beta^{ij}x^j\\
  &\equiv& \sum_{j=0}^{\delta-2}x^j\sum_{i=0}^{q^m-2}\frac{c_i}{g(\beta^{i})}\beta^{ij}
({\rm mod}x^{\delta-1}).
\end{eqnarray*}
Thus
\begin{eqnarray*}
  &&\sum_{i=0}^{q^m-2}\frac{c_i}{g(\beta^{i})}\frac{1}{1-\beta^ix}\equiv0({\rm mod }x^{\delta-1})\\
&\Leftrightarrow& \sum_{j=0}^{\delta-2}x^j\sum_{i=0}^{q^m-2}\frac{c_i}{g(\beta^{i})}\beta^{ij}\equiv0({\rm mod }x^{\delta-1})\\
&\Leftrightarrow& \sum_{i=0}^{q^m-2}\frac{c_i}{g(\beta^{i})}\beta^{ij}=0,\
j=0,\cdots,\delta-2\\
&\Leftrightarrow&c\in \Gamma(g,\delta,m,q),
\end{eqnarray*}
where the last equivalence follows from the last lemma.
The lemma is proved.
\end{proof}
\begin{lemma}[codeword equations]
$(c_o)_{o\in O(m,q)}\in{\rm R}\Gamma(g,\delta,m,q)$
if and only if $$\sum_{o\in O(m,q)}c_o\sum_{i\in o}\frac{\beta^{ij}}{g(\beta^i)}=0,\ j=0,1,\cdots,\delta-2,$$ if and only if
$$\sum_{o\in O(m,q)}c_o\sum_{i\in o}\frac{1}{g(\beta^{i})(1-\beta^i x)}\equiv0 ({\rm mod} x^{\delta-1}).$$
\end{lemma}
\begin{proof} Let
$(c_o)_{o\in O(m,q)}\in{\rm GF}(q)^{O(m,q)}$. For $i\in{\mathbb Z}/(q^m-1)$,
define
$$c_i=c_o, \ i\in o.$$ Then $$\sum_{i=0}^{q^m-2}\frac{c_i}{g(\beta^i)}\beta^{ij}=\sum_{o\in O(m,q)}c_o\sum_{i\in o}\frac{\beta^{ij}}{g(\beta^i)},\ j=0,1,\cdots,\delta-2.$$
It follows that
\begin{eqnarray*}
  &&(c_o)_{o\in O(m,q)}\in{\rm R}\Gamma(g,\delta,m,q)\\
&\Leftrightarrow&
 (c_i)_{i\in{\mathbb Z}/(q^m-1)}\in \Gamma^{\rm F}(g,\delta,m,q)\\
   &\Leftrightarrow& \sum_{i=0}^{q^m-2}\frac{c_i}{g(\beta^i)}\beta^{ij}=0,\ j=0,1,\cdots,\delta-2\\
&\Leftrightarrow&\sum_{o\in O(m,q)}c_o\sum_{i\in o}\frac{\beta^{ij}}{g(\beta^i)}=0,\ j=0,1,\cdots,\delta-2.
\end{eqnarray*}
Similarly, $c\in \Gamma(g,\delta,m,q)$ if and only if
$$\sum_{i=0}^{q^m-2}\frac{c_i}{g(\beta^{i})}\frac{1}{1-\beta^ix}\equiv0({\rm mod }x^{\delta-1}).$$
The lemma is proved.
\end{proof}\begin{theorem}[dimension bound]\label{dimension-bound}$${\rm dim}{\rm R}\Gamma(g,\delta,m,q)\geq |O(m,q)|-\delta+1.$$
\end{theorem}
\begin{proof} This follows from the last lemma as well as the fact that
$$\sum_{i\in o}\frac{\beta^{ij}}{g(\beta^i)}\in{\rm GF}(q).$$\end{proof}
\section{The Gilbert-Varshamov bound}
\hskip 0.1in In this section we prove Theorem \ref{GV-bound}, which says that the reduced Goppa codes approach the Gilbert-Varshamov bound as  classical Goppa codes \cite{Be73}.
\begin{definition}[Entropy]The $q$-ary entropy function on $(0,1)$ is defined by the formula
$$H_q(x)=x\log_q(q-1)-x\log_qx-(1-x)\log_q(1-x).$$\end{definition}
\begin{lemma}[Entropy lemma]If
$\frac{d}{n}\leq\frac{q-1}{q}$, then
$$H_q(\frac{d}{n})\geq \frac{1}{n}\log_q\sum_{i=0}^d{n\choose i}(q-1)^{i}.$$\end{lemma}
\begin{proof} By assumption, we have $\frac{d}{n-d}\leq q-1.$ Thus, if $0\leq i\leq d$, we have
\begin{eqnarray*}
   q^{-nH_q(\frac{d}{n})}&=& (q-1)^{-d}(\frac{d}{n})^d(\frac{n-d}{n})^{n-d} \\
   &\leq& (q-1)^{-i}(\frac{d}{n})^i(\frac{n-d}{n})^{n-i}\\
\end{eqnarray*}
It follows that
\begin{eqnarray*}
   q^{-nH_q(\frac{d}{n})}\sum_{i=0}^d{n\choose i}(q-1)^{i}
   &\leq& \sum_{i=0}^d{n\choose i}(\frac{d}{n})^i(\frac{n-d}{n})^{n-i}\\
   &\leq&1.
\end{eqnarray*}
The lemma now follows.
\end{proof}\begin{definition}The set of monic irreducible polynomials over ${\rm GF}(q)$ is denoted as
${\rm Irr}(\delta-1,q)$.\end{definition}
\begin{lemma}
\begin{eqnarray*}%                                  numbering (before each equation)
 && \#\{g\in {\rm Irr}(\delta-1,q)\mid {\rm dist}({\rm R}\Gamma(g,\delta,m,q))\leq d\} \\
 &\leq & \frac{md}{\delta-1}\sum_{i=0}^d{|O(m,q)|\choose i}(q-1)^i.
\end{eqnarray*}\end{lemma}

\begin{proof} We have
\begin{eqnarray*}
  &&\#\{g\in {\rm Irr}(\delta-1,q)\mid c\in {\rm R}\Gamma(g,\delta,m,q)\}\\
  &\leq&\#\{g\in {\rm Irr}(\delta-1,q)\mid \sum_{i=0}^{q^m-2}\frac{c_i}{x-\beta^i}\equiv0({\rm mod} g(x))\}\\
  &\leq &\frac{m\cdot{\rm wt}(c)}{\delta-1}.
\end{eqnarray*}
It follows that
\begin{eqnarray*}%                                  numbering (before each equation)
 && \#\{g\in {\rm Irr}(\delta-1,q)\mid {\rm dist}({\rm R}\Gamma(g,\delta,m,q))\leq d\}\\
&\leq & \sum_{i=1}^d\sum_{{\rm wt}(c)=i}\#\{g\in {\rm Irr}(\delta-1,q)\mid c\in {\rm R}\Gamma(g,\delta,m,q)\}\\
 &\leq & \frac{md}{\delta-1}\sum_{i=0}^d{|O(m,q)|\choose i}(q-1)^i.
\end{eqnarray*}
The lemma is proved.
\end{proof}
\begin{theorem}[Asymptotic Gibert-Varshamov bound]\label{GV-bound}Let $q$ be any prime power. Let $\varepsilon>0$. Then there is a tuple $(g,\delta,m)$ such that ${\rm R}\Gamma(g,\delta,m,q)$ makes sense, such that
 $$\frac{{\rm dist}{\rm R}\Gamma(g,\delta,m,q)}{|O(m,q)|}\leq \frac{q-1}{q},$$
and such that
$$H_q(\frac{{\rm dist}{\rm R}\Gamma(g,\delta,m,q)}{|O(m,q)|})+\frac{{\rm dim}{\rm R}\Gamma(g,\delta,m,q)}{|O(m,q)|}\geq 1-\varepsilon.$$\end{theorem}
\begin{proof}
First, we choose $N>0$ such that, for all $m>N$, and for all
$\delta$ with $$\frac{q^m}{m^2}\leq\delta\leq\frac{q^m}{m}+q^{\frac{m}{2}+1},$$
we have
$$\frac{\log_q(1-q^{-\frac{\delta-1}{2}}-q^{-\delta+1})
+\log_q(\delta-1)-m-\log_qm}{|O(m,q)|}
\geq-\varepsilon.$$
Second, we choose $m$ and
$\delta$ such that $m>N$, such that $\delta-1$ is prime to $m$,
and such that
$$\frac{q^m}{m^2}\leq\delta\leq\frac{q-1}{q}\cdot|O(m,q)|.$$
Third, we choose $d$ to be the largest integer such that
               $$\frac{md}{\delta-1}\sum_{i=0}^d{|O(m,q)|\choose i}(q-1)^i<q^{\delta-1}-q^{\frac{\delta-1}{2}}-1,$$which implies that
$$\#\{g\in {\rm Irr}(\delta-1,q)\mid {\rm dist}({\rm R}\Gamma(g,\delta,m,q))\leq d\}<q^{\delta-1}-q^{\frac{\delta-1}{2}}-1<
\# {\rm Irr}(\delta-1,q).$$
Fourth, we choose $g\in{\rm Irr}(\delta-1,q)$ such that
$$g\not\in\{g\in {\rm Irr}(\delta-1,q)\mid {\rm dist}({\rm R}\Gamma(g,\delta,m,q))\leq d\}.$$
As $\delta-1$ is prime to $m$, $g$ has no roots in ${\rm GF}(q^m)^{\times}$.
Thus ${\rm R}\Gamma(g,\delta,m,q)$ makes sense, and
               $$D={\rm dist}({\rm R}\Gamma(g,\delta,m,q))> d.$$

With these chosen parameters, we have
\begin{eqnarray*}
  \frac{{\rm dist}({\rm R}\Gamma(g,\delta,m,q))}{|O(m,q)|} &\leq & \frac{|O(m,q)|-{\rm dim}{\rm R}\Gamma(g,\delta,m,q)+1}{|O(m,q)|} \\
  &\leq & \frac{\delta}{|O(m,q)|}\\
&\leq&\frac{q-1}{q},
\end{eqnarray*}
where the first inequality follows from the singleton bound, and the second inequality follows from our bound for the dimensions of reduced Goppa codes.
Moreover,
\begin{eqnarray*}
H_q(\frac{{\rm dist}{\rm R}\Gamma(g,\delta,m,q)}{|O(m,q)|}) &\geq& \frac{1}{|O(m,q)|}\log_q\sum_{i=0}^D{|O(m,q)|\choose i}(q-1)^i
\\
&\geq&\frac{\log_q(q^{\delta-1}-q^{\frac{\delta-1}{2}}-1)+\log_q(\delta-1)-\log_q(mD)}{|O(m,q)|}
 \\
&\geq&\frac{\log_q(q^{\delta-1}-q^{\frac{\delta-1}{2}}-1)
+\log_q(\delta-1)-m-\log_q(m)}{|O(m,q)|}\\
  &\geq&\frac{\delta-1}{|O(m,q)|}-\varepsilon\\
  &\geq&\frac{|O(m,q)|-{\rm dim}{\rm R}\Gamma(g,\delta,m,q)}{|O(m,q)|}-\varepsilon.
             \end{eqnarray*}
The theorem is proved.
\end{proof}
\section{Decoding}
\hskip 0.1in In this section we describe the decoding algorithm for reduced Goppa codes.
The main result is Theorem \ref{decoding-time}.

\begin{definition}Let $r\in{\rm GF}(q)^{O(m,q)}$.
If there is a codeword $c\in{\rm R}\Gamma(g,\delta,m,q)$ such that
$$\sum_{o\in O(m,q):r_o\neq c_o}|o|\leq \frac{\delta-1}{2},$$ then it is called a B-M  correctable word for ${\rm R}\Gamma(g,\delta,m,q)$.\end{definition}
\begin{definition}If $r\in{\rm GF}(q)^{O(m,q)}$
is a B-M  correctable word for ${\rm R}\Gamma(g,\delta,m,q)$, then the error location set of $r$ is $$E=\{o\in O(m,q)\mid r_o\neq c_o\},$$ and the error of $r$
is $e=r-c$.\end{definition}\begin{definition}If $r\in{\rm GF}(q)^{O(m,q)}$
is a B-M  correctable word for ${\rm R}\Gamma(g,\delta,m,q)$, then
  $$S_j=\sum_{o\in O(m,q)}r_o\sum_{i\in o}\frac{\beta^{ij}}{g(\beta^i)},\ j=0,1,\cdots,\delta-2.$$
   are called syndromes of $r$.
\end{definition}
\begin{lemma}\label{syndrome-time}If $r\in{\rm GF}(q)^{O(m,q)}$
is a B-M  correctable word for ${\rm R}\Gamma(g,\delta,m,q)$, then
its syndromes can be calculated out within
  $O(n\delta)$ time in $q$-ary field operations, where $n$ is the length of $r$.
\end{lemma}
\begin{proof} This follows from the fact that
$$\sum_{i\in o}\frac{\beta^{ij}}{g(\beta^i)}\in{\rm GF}(q),\ j=0,1,\cdots,\delta-2,$$
which can be calculated out in pre-computation.
\end{proof}
  \begin{lemma}[syndrome congruence]If $r\in{\rm GF}(q)^{O(m,q)}$
is a B-M  correctable word for ${\rm R}\Gamma(g,\delta,m,q)$, then
  $$\sum_{j=0}^{\delta-2}S_jx^j\equiv\sum_{o\in O(m,q)}e_o\sum_{i\in o}\frac{1}{g(\beta^{i})(1-\beta^i x)}({\rm mod }x^{\delta-1}).$$
  \end{lemma}
\begin{proof} We have
\begin{eqnarray*}
  \sum_{j=0}^{\delta-2}S_jx^j &=& \sum_{j=0}^{\delta-2}x^j\sum_{o\in O(m,q)}r_o\sum_{i\in o}\frac{\beta^{ij}}{g(\beta^i)} \\
  &=&\sum_{o\in O(m,q)}r_o\sum_{i\in o}\frac{\sum_{j=0}^{\delta-2}x^j\beta^{ij}}{g(\beta^i)}\\
&\equiv&\sum_{o\in O(m,q)}r_o\sum_{i\in o}\frac{1}{g(\beta^{i})(1-\beta^i x)} ({\rm mod} x^{\delta-1}).\end{eqnarray*}
The lemma now follows from the equality $r=c+e$ and the codeword equation
$$\sum_{o\in O(m,q)}c_o\sum_{i\in o}\frac{1}{g(\beta^{i})(1-\beta^i x)}\equiv0 ({\rm mod} x^{\delta-1}).$$ \end{proof}
\begin{definition}If $r\in{\rm GF}(q)^{O(m,q)}$
is a B-M  correctable word for ${\rm R}\Gamma(g,\delta,m,q)$, then the polynomial pair $(\omega(x),\sigma(x))$  such that $\deg\omega(x)<\deg\sigma(x)\leq\frac{\delta-1}{2}$, and
$$\frac{\omega(x)}{\sigma(x)}\equiv\sum_{j=0}^{\delta-2}S_jx^j({\rm mod }x^{\delta-1})$$
is called a B-M pair of $r$.
\end{definition}
\begin{lemma}\label{BM-time}If $r\in{\rm GF}(q)^{O(m,q)}$
is a B-M  correctable word for ${\rm R}\Gamma(g,\delta,m,q)$, then, after its syndromes have been calculated out, its B-M pair $(\omega(x),\sigma(x))$  can be calculated out within
  $O(\delta^2)$ time in $q$-ary field operations.
\end{lemma}
\begin{proof} This follows from the time estimation for the well-known Berlekamp-Massey algorithm, as well as the fact that
$$S_j\in{\rm GF}(q),\ j=0,\cdots,\delta-2.$$ \end{proof}
  \begin{definition}The location polynomial of $o\in O(m,q)$ is the irreducible polynomial
$$p_o(x)=\prod_{i\in o}(x-\beta^{-i}).$$
  \end{definition}
  \begin{lemma}[syndrome equation]If $r\in{\rm GF}(q)^{O(m,q)}$
is a B-M  correctable word for ${\rm R}\Gamma(g,\delta,m,q)$, then$$\frac{\omega(x)}{\sigma(x)}=\sum_{o\in O(m,q)}e_o\sum_{i\in o}\frac{1}{g(\beta^{i})(1-\beta^i x)}.$$
  \end{lemma}
  \begin{proof} By the syndrome congruence, as well as the definition of B-M pair, we have
\begin{eqnarray*}
  \frac{\omega(x)}{\sigma(x)} &\equiv& \sum_{j=0}^{\delta-2}S_jx^j({\rm mod }x^{\delta-1}) \\
  &\equiv& \sum_{o\in O(m,q)}e_o\sum_{i\in o}\frac{1}{g(\beta^{i})(1-\beta^i x)}({\rm mod }x^{\delta-1}).
\end{eqnarray*}Comparing degrees of related polynomials, we see that
  $$\frac{\omega(x)}{\sigma(x)}=\sum_{o\in O(m,q)}e_o\sum_{i\in o}\frac{1}{g(\beta^{i})(1-\beta^i x)}.$$
The lemma is proved.
  \end{proof}
  \begin{lemma}[error locating formula]If $r\in{\rm GF}(q)^{O(m,q)}$
is a B-M  correctable word for ${\rm R}\Gamma(g,\delta,m,q)$, then
$$E=\{o\in O(m,q):\ p_o\mid\sigma(x)\}.$$
  \end{lemma}
  \begin{proof} This follows from the last lemma.
  \end{proof}

Consulting historical results on the factorisation of polynomials over finite fields \cite{Be67, Be68, Be70}, \cite{Mc69}, \cite{Ca83}, \cite{MOV}, \cite{Ni93a, Ni93b}, \cite{RZ}, \cite{GS}, we introduce the following definition.
\begin{definition}Let  $q$ is a power of a small prime number. Let $\nu$ be the least nonnegative integer such that
every polynomial of degree $n$ over ${\rm GF}(q)$ can be factored into irreducible polynomials  ${\rm GF}(q)$ within $O(n^2(\log (nq))^{\nu})$ time in $q$-ary field operations.\end{definition}
  \begin{lemma}[error locating time]\label{locating-time}Let  $q$ is a power of a small prime number. If $r$
is a B-M  correctable word for ${\rm R}\Gamma(g,\delta,m,q)$, then, after its B-M pair has been calculated out,
its error location set $E$ can be calculated out within $O(\delta^2(\log \delta)^{\nu})$
time in $q$-ary field operations.
  \end{lemma}
  \begin{proof} This follows from the last lemma, and the fact that $\sigma(x)$ is a polynomial over ${\rm GF}(q)$.
  \end{proof}
  \begin{lemma}[error correction formula]If $r\in{\rm GF}(q)^{O(m,q)}$
is a B-M  correctable word for ${\rm R}\Gamma(g,\delta,m,q)$, and $o\in E$, then
$$e_o\equiv-\frac{\omega(x)g(1/x)}{x\sigma'(x)}({\rm mod }p_o(x)).$$
  \end{lemma}
  \begin{proof} In fact, for $i\in o$, we have
  \begin{eqnarray*}
    \omega(x) &\equiv&-e_o\frac{\beta^{-i}\sigma(x)}{g(\beta^{i})(x-\beta^{-i})}({\rm mod} (x-\beta^{-i})) \\
   &\equiv&-e_o\frac{x\sigma'(x)}{g(1/x)}({\rm mod} (x-\beta^{-i})).
  \end{eqnarray*}
Thus, by Chinese remainder theorem,
$$\omega(x)\equiv-e_o\frac{x\sigma'(x)}{g(1/x)}({\rm mod} p_o(x)).$$
The lemma now follows.
  \end{proof}
  \begin{lemma}[error correction time]\label{correction-time}If $r\in{\rm GF}(q)^{O(m,q)}$
is a B-M  correctable word for ${\rm R}\Gamma(g,\delta,m,q)$, then,
after its error location set $E$ has calculated out, its error can be corrected within $O(\delta^2)$
time in $q$-ary field operations.
  \end{lemma}
  \begin{proof} This follows from the last lemma.
  \end{proof}
\begin{theorem}[decoding time]\label{decoding-time}Let  $q$ is a power of a small prime. If $r$
is a B-M  correctable word for ${\rm R}\Gamma(g,\delta,m,q)$, and $n=|O(m,q)|$, then
$r$ can be decoded within
$O(n^2(\log n)^{\nu})$ time in $q$-ary field operations.
\end{theorem}
\begin{proof}
Theorem \ref{decoding-time} follows form Lemmas \ref{syndrome-time}, \ref{BM-time}, \ref{locating-time}, and \ref{correction-time}.
\end{proof}
Recall that, if $r$ is a received word of $\Gamma(g,\delta,m,q)$ and there is a codeword $c$ such that
$${\rm dist}(r,c)\leq \frac{\delta-1}{2},$$
then $r$ can be decoded within
$O(n^2(\log n)^{\nu})$ time in $q^m$-ary field operations with $n=q^m-1$. By comparison, reduced Goppa codes can be decoded much quicker that the classical Goppa codes.
\section{Comparison with Goppa codes}
\hskip 0.1in
In this section we compare the $m$-fold direct product
${\rm R}\Gamma(g,\delta,m,q)^m$ with the classical Goppa code $\Gamma(g,\delta,m,q)$. We illustrate their performances in the following table.
\begin{tabular}{|c|c|c|}
  \hline
   & ${\rm R}\Gamma(g,\delta,m,q)^m$ & $\Gamma(g,\delta,m,q)$ \\
  \hline
  codeword length & $q^m-1\leq n \leq q^m+mq^{\frac{m}{2}+1}$ & $n=q^m-1$ \\
  \hline
dimension & $k\geq n-m(\delta-1)$& $k\geq n-m(\delta-1)$ \\
  \hline
  minimal distance & $d\geq\delta$ & $d\geq\delta$ \\
  \hline
  Asymptotic & meets & meets \\
  Gilbert-Varshamov bound &  &  \\
  \hline
  decoding time & $O(n^2(\log n)^{\nu-1})$ & $O(n^2(\log n)^{\nu})$\\
   & 
in $q$-ary field operations & in $q^m$-ary field operations \\
  \hline
\end{tabular}

\end{document}